\documentclass[%
 reprint,
nofootinbib,
nobibnotes,
bibnotes,
 amsmath,amssymb,
 aps,
pra,
]{revtex4-1}

\usepackage{graphicx}
\usepackage{dcolumn}
\usepackage{bm}

\begin{document}

\preprint{APS/123-QED}

\title{Charge Fractionalization in Oxide Heterostructures: A Field Theoretical Model}

\author{M. Karthick Selvan}
\email{karthick.selvan@yahoo.com}
\author{Prasanta K. Panigrahi}%
 \email{pprasanta@iiserkol.ac.in}
\affiliation{
 Department of Physical Sciences, Indian Institute of Science Education and Research Kolkata, Mohanpur, West Bengal 741246, India\\
}%





\begin{abstract}
LaAlO$_{3}$/SrTiO$_{3}$ heterostructure with polar and non-polar constituents has been shown to exhibit interface metallic conductivity due to fractional charge transfer to the interface. The interface reconstruction by electron redistribution along (001) orientation, in which half of an electron is transferred per two dimensional unit cell to the adjacent planes, resulting in a net transfer of half of the charge, to both the interface and top most atomic planes, has been ascribed to a polar discontinuity at the interface in the Polar Catastrophe model. This avoids the divergence of electrostatic potential, as the number of layers are increased, producing an oscillatory electric field and finite potential. Akin to the description of charge fractionalization in quasi one-dimensional polyacetylene by the field theoretic Jackiw-Rebbi model with fermions interacting with topologically non-trivial background field, we show an analogous connection between polar catastrophe model and Bell-Rajaraman model, where the charge fractionalization occurs in the soliton free sector as an end effect.  
\end{abstract}

\pacs{Valid PACS appear here}
\maketitle


\section{Introduction}
The quasi-two dimensional electron gas formed at the interface of LaAlO$_{3}$/SrTiO$_{3}$ heterostructure has been shown to have interesting properties such as metallic conductivity \cite{ohtomo2004high,thiel2006tunable,caviglia2008electric,bell2009dominant}, ferromagnetism \cite{lee2013titanium,brinkman2007magnetic}, superconductivity \cite{ueno2008electric} and the coexistence of magnetic order and superconductivity \cite{bert2011direct}. This heterostructure consists of LaAlO$_{3}$ film deposited layer by layer over SrTiO$_{3}$ substrate along the (001) orientation. LaAlO$_{3}$ has alternating positively charged (LaO)$^+$ and negatively charged (AlO$_{2}$)$^-$ atomic planes along the (001) orientation. The electric field that arises between these oppositely charged atomic planes gives rise to an in-built electrostatic potential that diverges with the thickness. When this polar LaAlO$_{3}$ is deposited over non-polar SrTiO$_{3}$ substrate, which consists of alternating neutral (SrO)$^0$ and (TiO$_2$)$^0$ atomic planes, the polar discontinuity that arises at the interface, causes this heterostructure to be electrostatically unstable. The stable LaAlO$_{3}$/SrTiO$_{3}$ heterostructure requires interface reconstruction which could be ionic reconstruction involving lattice distortion or cationic inter-mixing. Polar catastrophe model, a simple electrostatic model, was proposed to explain the formation of two dimensional electron gas at the interface of LaAlO$_{3}$/SrTiO$_{3}$ heterostructure \cite{ohtomo2004high,nakagawa2006some}. This model proposes that the interface is reconstructed by electron redistribution in which one half of an electron or a hole is transferred per two dimensional unit cell along the (001) orientation depending on the interface structure. This charge transfer results in a net transfer of half of an electron or a hole to the interface. Although this model does not explain many effects, such as strong polar distortions that delay the onset of metallic conductivity \cite{pentcheva2009avoiding} and the reversible metal-insulator transition upon external bias \cite{caviglia2008electric,cen2008nanoscale}, this gives a possible explanation for the formation of two dimensional electron gas at the interface of LaAlO$_{3}$/SrTiO$_{3}$ heterostructure.

In relativistic quantum field theory, charge fractionalization has been shown to occur by Jackiw and Rebbi (JR) \cite{jackiw1976solitons}, in the case of spinless fermions interacting with topologically non-trivial background field. The phenomenon of charge fractionalization has been successfully used by SSH model to explain the observed conductivity in the quasi-one dimensional polyacetylene \cite{su1979solitons,rice1979charged,su1980soliton,rao2008fermion}. The similarity between these two different models has been established by Jackiw and Schrieffer \cite{jackiw1981solitons}. Both models give rise to spontaneous breaking of the reflection symmetry, which results in the formation of doubly degenerate ground states. There exists soliton excitations which interpolate between these two degenerate ground states. In both cases, the charge conjugation symmetry assures the existence of unpaired zero energy state. The charge is $+\frac{1}{2}$ if the zero energy state is occupied, otherwise it is $-\frac{1}{2}$. Bell and Rajaraman \cite{bell1983states}, in studying the system of discrete lattice sites related to the polyacetylene model, have shown that fermion number fractionalization can occur in the soliton free sector, as an end effect.

In this paper, we demonstrate the analogy between the polar catastrophe model which explains the formation of two dimensional electron gas at the interface of LaAlO$_{3}$/SrTiO$_{3}$ heterostructure and the Bell-Rajaraman model where charge fractionalization occurs as an end effect. By this, we show the first physical system where Bell-Rajaraman model can be applied to explain the observed phenomenon.   
\section{Polar Catastrophe Model (PCM)}
ABO$_{3}$ perovskites have alternating AO and BO$_{2}$ atomic planes along the (001) orientation. SrTiO$_{3}$ is of A$^{2+}$B$^{4+}$O$_{3}$ type and has alternating neutral (SrO)$^0$ and (TiO$_2$)$^0$ atomic planes. However, LaAlO$_{3}$ is of A$^{3+}$B$^{3+}$O$_{3}$ type, which has alternating (LaO)$^{+}$ and (AlO$_{2}$)$^{-}$ atomic planes. Polar LaAlO$_{3}$ has an in-built electrostatic potential that diverges with the thickness. Polar discontinuity that arises at the interface of this polar and non-polar heterostructure makes it electrostatically unstable and the stable structure requires interface reconstruction. Since cations in the perovskite structures can have mixed valance states, energetically the electron redistribution along the (001) orientation is favorable than any other form of interface reconstruction. LaAlO$_{3}$/SrTiO$_{3}$ heterostructure can have two different interface structures: (LaO)$^{+}$/(TiO$_2$)$^{0}$ (FIG.1a) and (AlO$_{2}$)$^{-}$/(SrO)$^{0}$ (FIG.1b). 
\begin{figure}
\centering
\includegraphics[width=8.5cm, height=8.5cm]{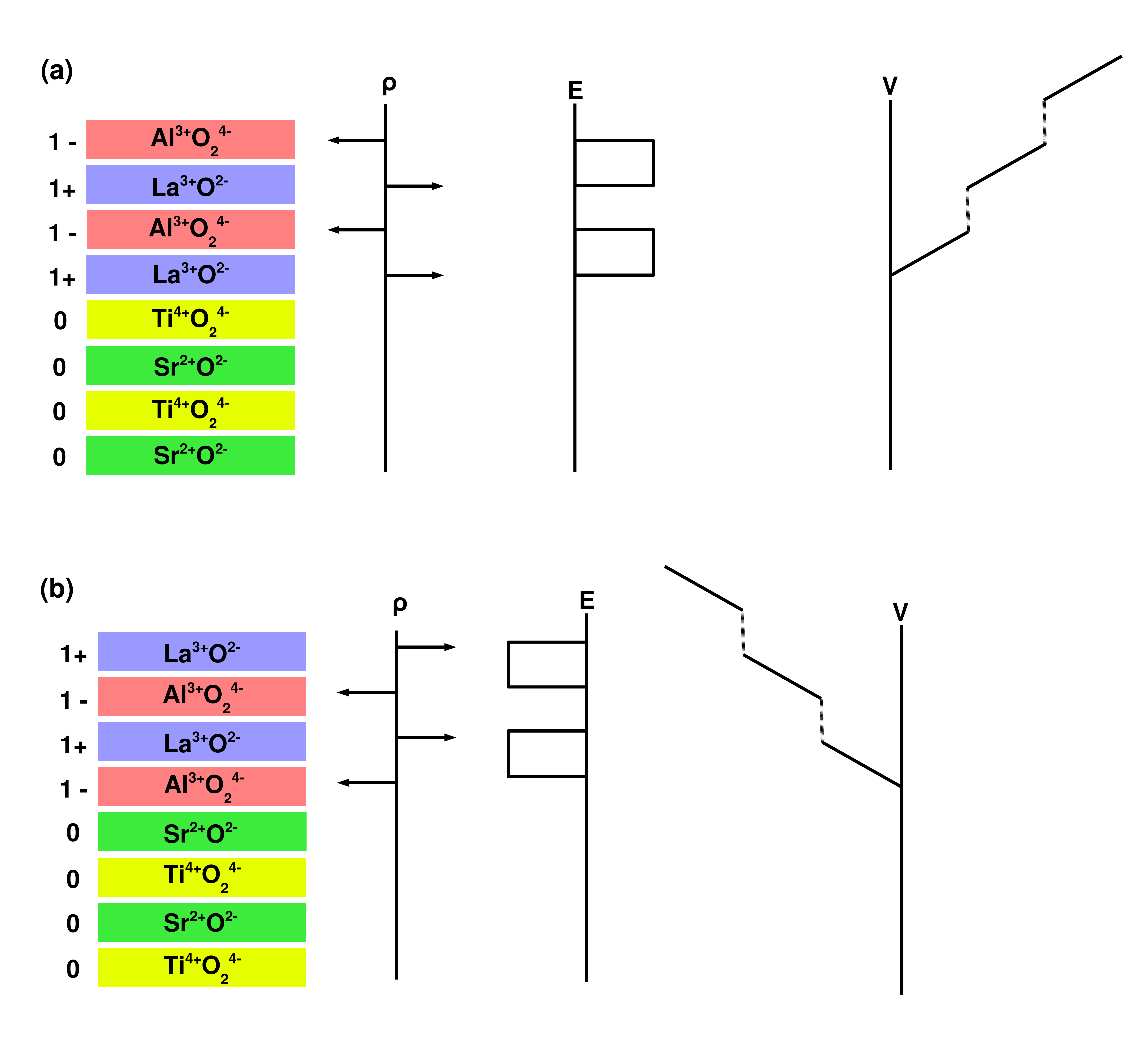}
\caption{(001) oriented LaAlO$_{3}$/SrTiO$_{3}$ heterostructure. (a) Unreconstructed (LaO)$^{+}$/(TiO$_{2}$)$^{0}$ interface. SrTiO$_{3}$ has alternating neutral atomic planes and LaAlO$_{3}$ has alternatively charged polar planes. The non-zero electric field between the oppositely charged atomic planes of LaAlO$_{3}$ creates an in-built electrostatic potential that diverges with the thickness of LaAlO$_{3}$. (b) Unreconstructed (AlO$_{2}$)$^{-}$/(SrO)$^{0}$ interface. In this case, the electrostatic potential diverges negatively with the thickness.}
\end{figure}

According to polar catastrophe model, one half of an electron (or a hole) is transferred per two dimensional unit cell in (LaO)$^{+}$/(TiO$_2$)$^{0}$ (or (AlO$_{2}$)$^{-}$/(SrO)$^{0}$) interface, hence it is of n-type (or p-type) interface. This results in a net transfer of half electron or hole to the neutral layer at the interface and to the top most layer of LaAlO$_{3}$ and produces an interface dipole that causes the electric field to oscillate about zero and potential to remain finite (FIG.2).  
\begin{figure}
\centering
\includegraphics[width=8.5cm, height=8.5cm]{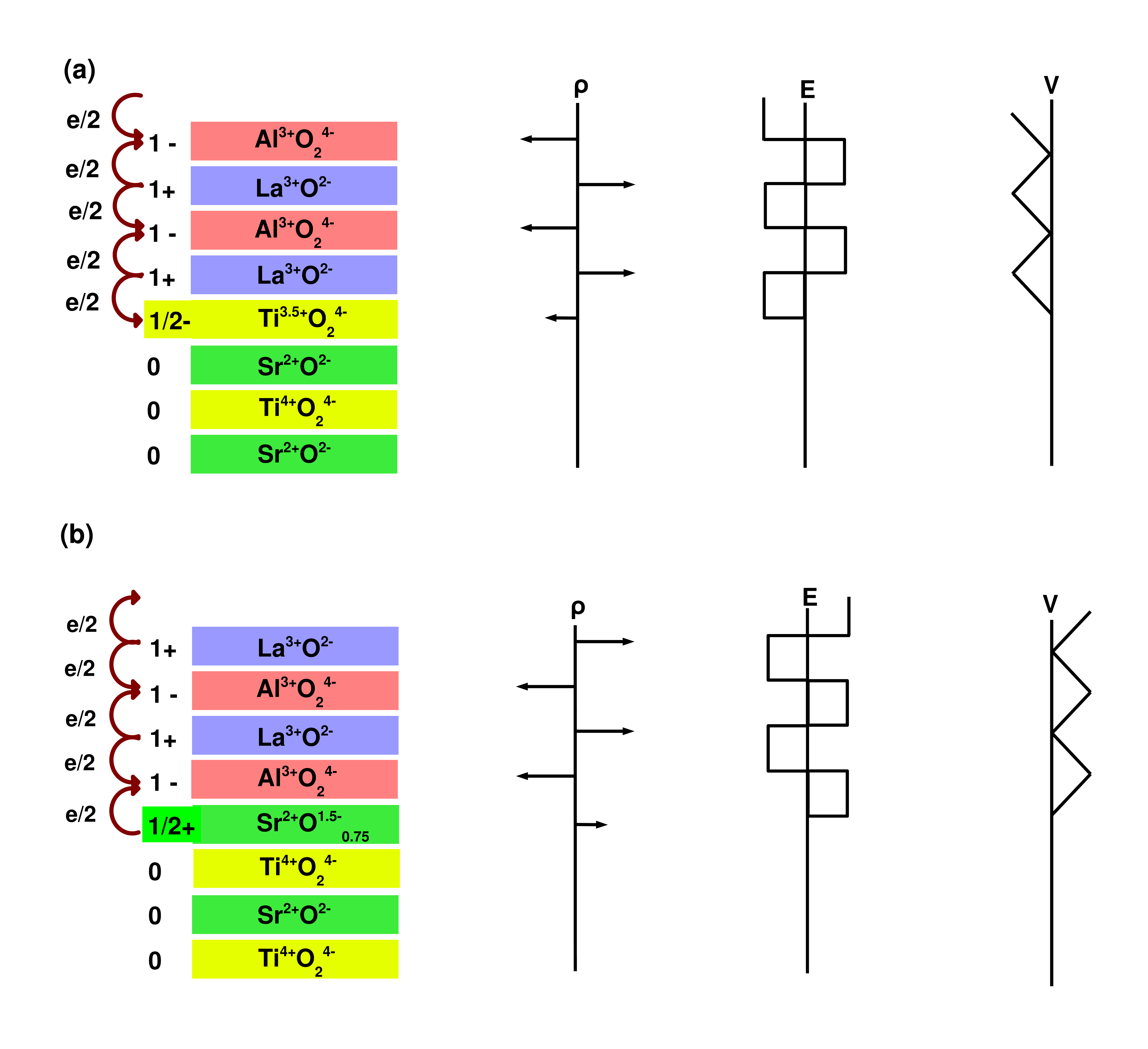}
\caption{Polar discontinuity at the interface forces electron redistribution along the (001) orientation. (a) Half an electron per two dimensional unit cell is transferred in the case of (LaO)$^{+}$/(TiO$_{2}$)$^{0}$ interface. This adds a net half electron to the neutral TiO$_{2}$ plane near the interface and to the top most AlO$_{2}$ plane (not shown here). (b) In (AlO$_{2}$)$^{-}$/(SrO)$^{0}$ interface, half a hole per two dimensional unit cell is transferred. In both cases, the charge redistribution produces an interface dipole that changes the electric field to oscillate about zero and the potential to remain finite.}
\end{figure}

However, experimentally only the n-type interface, (LaO)$^{+}$/(TiO$_2$)$^{0}$, was found conducting and the p-type interface, (AlO$_{2}$)$^{-}$/(SrO)$^{0}$, remained insulating. Since Ti cation can exist in Ti$^{3+}$ and Ti$^{4+}$ valance states, in the n-type interface, a net half electron is transferred from LaAlO$_{3}$ into the Ti 3d conduction band of SrTiO$_{3}$, close to the interface that gives rise to conduction. It was experimentally proved that oxygen vacancy and cation inter-mixing do not contribute to the interface reconstruction and hence to the conductivity observed in the n-type interface \cite{ohtomo2004high,cantoni2012electron}. However, there exists a threshold thickness (3 unit cells) of LaAlO$_{3}$, above which the metallic conductivity was observed \cite{cantoni2012electron,thiel2006tunable}. From optical studies, it was shown that up to the thickness of 3 unit cells, the transferred electrons are trapped in the localized states of interface created by lattice distortion \cite{savoia2009polar,asmara2014mechanisms}. Band structure studies imply that the lattice distortion due to ionic relaxations are more at the outer layers of LaAlO$_{3}$ film that are adjacent to the SrTiO$_{3}$ and to the air; the distortion decreases and eventually vanishes as the thickness of LaAlO$_{3}$ is increased \cite{pauli2011evolution,salluzzo2013structural}. Further the electron carrier density measured by optical conductivity studies was found to agree with the value predicted by PCM \cite{savoia2009polar,asmara2014mechanisms}, which supports the validity of PCM.

In the case of p-type interface, a net hole transfer is necessary to avoid the potential divergence but the unavailability of mixed valance states to accommodate this half hole requires atomic reconstruction. It was experimentally shown that the p-type interface is reconstructed by oxygen vacancy, however, the oxygen vacancies which generally act as electron donor do not give rise to any conduction in the p-type interface \cite{nakagawa2006some}. 

\section{Comparision with Bell-Rajaraman Model}
A two dimensional view of (LaO)$^{+}$/(TiO$_{2}$)$^{0}$ interface is shown in FIG.3a. As the polar discontinuity at interface forces the electron redistribution along the (001) orientation, we consider only the inter-planar electron hopping along the z-direction. We use the Hamiltonian of polyacetylene model used in Ref. \cite{bell1983states}, which describes the nearest-neighbor electron hopping in one dimension and its interaction with the lattice displacement, that was taken as given. The displacement of atomic planes in the (LaO)$^{+}$/(TiO$_{2}$)$^{0}$ interface is considered as follows. The (LaO)$^{+}$ atomic planes are displaced by an amount $\alpha$ and the (AlO$_{2}$)$^{-}$ planes are displaced by an amount $\beta$. The inter-planar distance between the adjacent planes changes by $(-)^{n}{\delta} (={\alpha} + {\beta})$ (FIG. 3b).   \\
 \begin{figure}
\centering
\includegraphics[width=8.5cm, height=11.5cm]{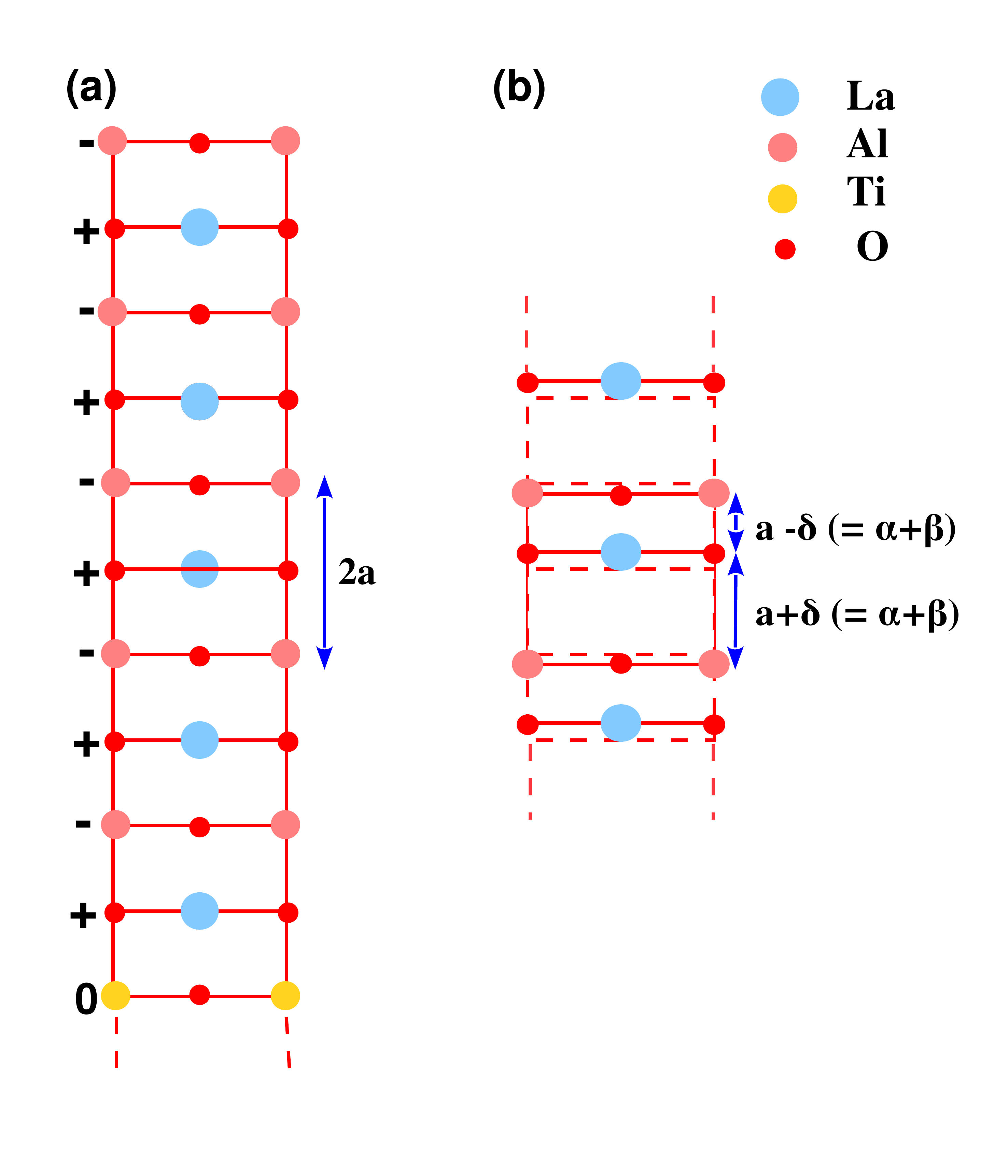}
\caption{(a) Two dimensional view of (LaO)$^{+}$/(TiO$_{2}$)$^{0}$ interface structure consisting of $2M$ alternatively charged (LaO)$^{+}$ and (AlO$_{2}$)$^{-}$ planes and a neutral (TiO$_{2}$)$^{0}$ plane. (b) Atomic planes are displaced from their initial position. (LaO)$^{+}$ plane is displaced by an amount ${\alpha}$ and (AlO$_{2}$)$^{0}$ planes are displaced by an amount ${\beta}$ (not specified here). The inter-planer distance changes by $(-)^{n}{\delta} (={\alpha} + {\beta})$ for adjacent planes.}
\end{figure}
 For this system, the Hamiltonian can be written in the form,
 \begin{equation}
\begin{aligned}H\,\,=\,\, & \underset{n}{\sum}\left[D_{n+1}^{\dagger}D_{n}+D_{n}^{\dagger}D_{n+1}\right]\left(\left(-\right)^{n}\delta-\frac{1}{2a}\right),\end{aligned}
\end{equation}
where $D{}_{n}$ and $D^{\dagger}{}_{n}$ are anti-commuting fermion creation and annihilation operators at the lattice point $n$:
 \begin{equation}
\begin{aligned}\left\{ D_{n},D_{m}^{\dagger}\right\} \,\,= & \,\,\delta_{nm}.\end{aligned}
\end{equation}

Using the index $n$ for even lattice sites and $m$ for odd lattice sites and defining the new operators:
\begin{equation}
\begin{aligned}C_{n}\,\,=\,\, & i^{n}D_{n}\,\,\,\,and\,\,\,\,B_{m}\,\,= \,\, i^{m+1}D_{m},\end{aligned}
\end{equation} 
the Hamiltonian can be rewritten in the form,
\[
\begin{aligned}H\,\,=\,\, & \underset{n}{\sum}\left[B_{n+1}^{\dagger}C_{n}+C_{n}^{\dagger}B_{n+1}\right]\left(\frac{1}{2a}-\delta\right)\end{aligned}
\]
\begin{equation}
\,\,\,\,\,\,\,\,\,\,\,\,\,\,\,\,\,\,\,\,\,\,\,\,\,\,\,\,\,\,\,\,\,\,\,\,-\underset{m}{\sum}\left[C_{m+1}^{\dagger}B_{m}+B_{m}^{\dagger}C_{m+1}\right]\left(\frac{1}{2a}+\delta\right).\end{equation}
The following commutation relations hold: 
\begin{equation}
\begin{aligned}\left[B_{m},H\right]\,\,=\,\, & \frac{1}{2a}\left(C_{m-1}-C_{m+1}\right)-\delta\left(C_{m-1}+C_{m+1}\right)\end{aligned}
\end{equation}

and
\begin{equation}
\begin{aligned}\left[C_{n},H\right]\,\,=\,\, & \frac{1}{2a}\left(B_{n+1}-B_{n-1}\right)-\delta\left(B_{n+1}+B_{n-1}\right).\end{aligned}
\end{equation}
From FIG.3, it is clear that only $2M$ atomic planes, corresponding to the alternating polar planes of LaAlO$_{3}$ and one neutral (TiO$_{2}$)$^{0}$ atomic plane that introduces the polar discontinuity at the interface, in total $2M+1$ atomic planes are of interest. $M$ can be either odd or even. Under the transformation,
 \begin{equation}
\begin{aligned}B_{m}\,\,=\,\,\sum_{r=-M}^{M} & b_{m}^{r}A_{r}\end{aligned}
\end{equation}
and
\begin{equation}
\begin{aligned}
C_{n}\,\,=\,\,\sum_{r=-M}^{M}c_{n}^{r}A_{r},\end{aligned}
\end{equation}

the Hamiltonian becomes,

\begin{equation}
\begin{aligned}H\,\,=\,\, & \sum_{r=-M}^{M}\varepsilon_{r}A_{r}^{\dagger}A_{r}.\end{aligned}
\end{equation}

 The orthonormal functions $\left(b_{m},c_{n}\right)$ satisfy the following equations,

\begin{equation}
\begin{aligned}\varepsilon_{r}b_{m}^{r}\,\,=\,\, & \frac{1}{2a}\left(c_{m-1}^{r}-c_{m+1}^{r}\right)-\delta\left(c_{m-1}^{r}+c_{m+1}^{r}\right)\end{aligned}
\end{equation}
and
\begin{equation}
\begin{aligned}\varepsilon_{r}c_{n}^{r}\,\,=\,\, & \frac{1}{2a}\left(b_{n+1}^{r}-b_{n-1}^{r}\right)-\delta\left(b_{n-1}^{r}+b_{n+1}^{r}\right).\end{aligned}
\end{equation}

Following the arguments of Ref. \cite{bell1983states} the charge conjugation symmetry of the above two equations assures the existence of a zero energy state for the system of $2M+1$ atomic planes. Depending on whether $M$ is odd or even the zero energy states are localized either at atomic planes labeled by odd numbers or at atomic planes labeled by even numbers. Further, by neglecting spin degree of freedom, the charge operators at lattice site $m$ and $n$ can be defined as,
\begin{equation}
\begin{aligned}\rho\left(m\right)\,\,=\,\, & B_{m}^{\dagger}B_{m}-\frac{1}{2}\end{aligned}
\end{equation}
and
\begin{equation}
\begin{aligned}\rho\left(n\right)\,\,=\,\, & C_{n}^{\dagger}C_{n}-\frac{1}{2}.\end{aligned}
\end{equation}
As a consequence of the choice of $2M+1$ atomic planes, when zero energy and positive energy states are unoccupied the charge is $-\frac{1}{2}$ and when the zero energy state is occupied, it is $+\frac{1}{2}$. It was shown that these half integral charges are localized, where the zero energy functions are localized. This is completely analogous to the PCM, in which the transfer of  half charge per two dimensional unit cell results in a net transfer of half charge to the end atomic planes, which are (TiO$_{2}$)$^{0}$ neutral plane at the interface and the top (AlO$_{2}$)$^{-}$ plane.\\
\section{Conclusion} 
 In conclusion, we have compared the electron redistribution along the (001)
 orientation of LaAlO$_{3}$/SrTiO$_{3}$ heterostructure proposed by polar catastrophe model to the Bell-Rajaraman model. In PCM, half of an electron per two dimensional unit cell is transferred along (001) orientation, across $2M+1$ atomic layers of LaAlO$_{3}$/SrTiO$_{3}$ heterostructure. This adds net half electron to the neutral TiO$_2$ layer and top AlO$_2$ layer. This is completely equivalent to the electron fractionalization process discussed in Bell-Rajaraman model, where the half integral charge is due to the specific choice of $2M+1$ lattice sites and the conjugation symmetry of the Hamiltonian. In polyacetylene, the model system considered in Ref. \cite{bell1983states}, the number of lattice points is uncertain. It could be either odd or even. But in LaAlO$_{3}$/SrTiO$_{3}$ heterostructure, electrons are redistributed exactly across $2M+1$ atomic planes. This is the first application of Bell-Rajaraman model to the best of authors' knowledge.
\bibliography{reference}

\begin{thebibliography}{23}%
\makeatletter
\providecommand \@ifxundefined [1]{%
 \@ifx{#1\undefined}
}%
\providecommand \@ifnum [1]{%
 \ifnum #1\expandafter \@firstoftwo
 \else \expandafter \@secondoftwo
 \fi
}%
\providecommand \@ifx [1]{%
 \ifx #1\expandafter \@firstoftwo
 \else \expandafter \@secondoftwo
 \fi
}%
\providecommand \natexlab [1]{#1}%
\providecommand \enquote  [1]{``#1''}%
\providecommand \bibnamefont  [1]{#1}%
\providecommand \bibfnamefont [1]{#1}%
\providecommand \citenamefont [1]{#1}%
\providecommand \href@noop [0]{\@secondoftwo}%
\providecommand \href [0]{\begingroup \@sanitize@url \@href}%
\providecommand \@href[1]{\@@startlink{#1}\@@href}%
\providecommand \@@href[1]{\endgroup#1\@@endlink}%
\providecommand \@sanitize@url [0]{\catcode `\\12\catcode `\$12\catcode
  `\&12\catcode `\#12\catcode `\^12\catcode `\_12\catcode `\%12\relax}%
\providecommand \@@startlink[1]{}%
\providecommand \@@endlink[0]{}%
\providecommand \url  [0]{\begingroup\@sanitize@url \@url }%
\providecommand \@url [1]{\endgroup\@href {#1}{\urlprefix }}%
\providecommand \urlprefix  [0]{URL }%
\providecommand \Eprint [0]{\href }%
\providecommand \doibase [0]{http://dx.doi.org/}%
\providecommand \selectlanguage [0]{\@gobble}%
\providecommand \bibinfo  [0]{\@secondoftwo}%
\providecommand \bibfield  [0]{\@secondoftwo}%
\providecommand \translation [1]{[#1]}%
\providecommand \BibitemOpen [0]{}%
\providecommand \bibitemStop [0]{}%
\providecommand \bibitemNoStop [0]{.\EOS\space}%
\providecommand \EOS [0]{\spacefactor3000\relax}%
\providecommand \BibitemShut  [1]{\csname bibitem#1\endcsname}%
\let\auto@bib@innerbib\@empty
\bibitem [{\citenamefont {Ohtomo}\ and\ \citenamefont
  {Hwang}(2004)}]{ohtomo2004high}%
  \BibitemOpen
  \bibfield  {author} {\bibinfo {author} {\bibfnamefont {A.}~\bibnamefont
  {Ohtomo}}\ and\ \bibinfo {author} {\bibfnamefont {H.}~\bibnamefont {Hwang}},\
  }\href@noop {} {\bibfield  {journal} {\bibinfo  {journal} {Nature}\ }\textbf
  {\bibinfo {volume} {427}},\ \bibinfo {pages} {423} (\bibinfo {year}
  {2004})}\BibitemShut {NoStop}%
\bibitem [{\citenamefont {Thiel}\ \emph {et~al.}(2006)\citenamefont {Thiel},
  \citenamefont {Hammerl}, \citenamefont {Schmehl}, \citenamefont {Schneider},\
  and\ \citenamefont {Mannhart}}]{thiel2006tunable}%
  \BibitemOpen
  \bibfield  {author} {\bibinfo {author} {\bibfnamefont {S.}~\bibnamefont
  {Thiel}}, \bibinfo {author} {\bibfnamefont {G.}~\bibnamefont {Hammerl}},
  \bibinfo {author} {\bibfnamefont {A.}~\bibnamefont {Schmehl}}, \bibinfo
  {author} {\bibfnamefont {C.}~\bibnamefont {Schneider}}, \ and\ \bibinfo
  {author} {\bibfnamefont {J.}~\bibnamefont {Mannhart}},\ }\href@noop {}
  {\bibfield  {journal} {\bibinfo  {journal} {Science}\ }\textbf {\bibinfo
  {volume} {313}},\ \bibinfo {pages} {1942} (\bibinfo {year}
  {2006})}\BibitemShut {NoStop}%
\bibitem [{\citenamefont {Caviglia}\ \emph {et~al.}(2008)\citenamefont
  {Caviglia}, \citenamefont {Gariglio}, \citenamefont {Reyren}, \citenamefont
  {Jaccard}, \citenamefont {Schneider}, \citenamefont {Gabay}, \citenamefont
  {Thiel}, \citenamefont {Hammerl}, \citenamefont {Mannhart},\ and\
  \citenamefont {Triscone}}]{caviglia2008electric}%
  \BibitemOpen
  \bibfield  {author} {\bibinfo {author} {\bibfnamefont {A.}~\bibnamefont
  {Caviglia}}, \bibinfo {author} {\bibfnamefont {S.}~\bibnamefont {Gariglio}},
  \bibinfo {author} {\bibfnamefont {N.}~\bibnamefont {Reyren}}, \bibinfo
  {author} {\bibfnamefont {D.}~\bibnamefont {Jaccard}}, \bibinfo {author}
  {\bibfnamefont {T.}~\bibnamefont {Schneider}}, \bibinfo {author}
  {\bibfnamefont {M.}~\bibnamefont {Gabay}}, \bibinfo {author} {\bibfnamefont
  {S.}~\bibnamefont {Thiel}}, \bibinfo {author} {\bibfnamefont
  {G.}~\bibnamefont {Hammerl}}, \bibinfo {author} {\bibfnamefont
  {J.}~\bibnamefont {Mannhart}}, \ and\ \bibinfo {author} {\bibfnamefont
  {J.-M.}\ \bibnamefont {Triscone}},\ }\href@noop {} {\bibfield  {journal}
  {\bibinfo  {journal} {Nature}\ }\textbf {\bibinfo {volume} {456}},\ \bibinfo
  {pages} {624} (\bibinfo {year} {2008})}\BibitemShut {NoStop}%
\bibitem [{\citenamefont {Bell}\ \emph {et~al.}(2009)\citenamefont {Bell},
  \citenamefont {Harashima}, \citenamefont {Kozuka}, \citenamefont {Kim},
  \citenamefont {Kim}, \citenamefont {Hikita},\ and\ \citenamefont
  {Hwang}}]{bell2009dominant}%
  \BibitemOpen
  \bibfield  {author} {\bibinfo {author} {\bibfnamefont {C.}~\bibnamefont
  {Bell}}, \bibinfo {author} {\bibfnamefont {S.}~\bibnamefont {Harashima}},
  \bibinfo {author} {\bibfnamefont {Y.}~\bibnamefont {Kozuka}}, \bibinfo
  {author} {\bibfnamefont {M.}~\bibnamefont {Kim}}, \bibinfo {author}
  {\bibfnamefont {B.}~\bibnamefont {Kim}}, \bibinfo {author} {\bibfnamefont
  {Y.}~\bibnamefont {Hikita}}, \ and\ \bibinfo {author} {\bibfnamefont
  {H.}~\bibnamefont {Hwang}},\ }\href@noop {} {\bibfield  {journal} {\bibinfo
  {journal} {Physical review letters}\ }\textbf {\bibinfo {volume} {103}},\
  \bibinfo {pages} {226802} (\bibinfo {year} {2009})}\BibitemShut {NoStop}%
\bibitem [{\citenamefont {Lee}\ \emph {et~al.}(2013)\citenamefont {Lee},
  \citenamefont {Xie}, \citenamefont {Sato}, \citenamefont {Bell},
  \citenamefont {Hikita}, \citenamefont {Hwang},\ and\ \citenamefont
  {Kao}}]{lee2013titanium}%
  \BibitemOpen
  \bibfield  {author} {\bibinfo {author} {\bibfnamefont {J.-S.}\ \bibnamefont
  {Lee}}, \bibinfo {author} {\bibfnamefont {Y.}~\bibnamefont {Xie}}, \bibinfo
  {author} {\bibfnamefont {H.}~\bibnamefont {Sato}}, \bibinfo {author}
  {\bibfnamefont {C.}~\bibnamefont {Bell}}, \bibinfo {author} {\bibfnamefont
  {Y.}~\bibnamefont {Hikita}}, \bibinfo {author} {\bibfnamefont
  {H.}~\bibnamefont {Hwang}}, \ and\ \bibinfo {author} {\bibfnamefont {C.-C.}\
  \bibnamefont {Kao}},\ }\href@noop {} {\bibfield  {journal} {\bibinfo
  {journal} {Nature materials}\ }\textbf {\bibinfo {volume} {12}},\ \bibinfo
  {pages} {703} (\bibinfo {year} {2013})}\BibitemShut {NoStop}%
\bibitem [{\citenamefont {Brinkman}\ \emph {et~al.}(2007)\citenamefont
  {Brinkman}, \citenamefont {Huijben}, \citenamefont {Van~Zalk}, \citenamefont
  {Huijben}, \citenamefont {Zeitler}, \citenamefont {Maan}, \citenamefont
  {Van~der Wiel}, \citenamefont {Rijnders}, \citenamefont {Blank},\ and\
  \citenamefont {Hilgenkamp}}]{brinkman2007magnetic}%
  \BibitemOpen
  \bibfield  {author} {\bibinfo {author} {\bibfnamefont {A.}~\bibnamefont
  {Brinkman}}, \bibinfo {author} {\bibfnamefont {M.}~\bibnamefont {Huijben}},
  \bibinfo {author} {\bibfnamefont {M.}~\bibnamefont {Van~Zalk}}, \bibinfo
  {author} {\bibfnamefont {J.}~\bibnamefont {Huijben}}, \bibinfo {author}
  {\bibfnamefont {U.}~\bibnamefont {Zeitler}}, \bibinfo {author} {\bibfnamefont
  {J.}~\bibnamefont {Maan}}, \bibinfo {author} {\bibfnamefont {W.}~\bibnamefont
  {Van~der Wiel}}, \bibinfo {author} {\bibfnamefont {G.}~\bibnamefont
  {Rijnders}}, \bibinfo {author} {\bibfnamefont {D.}~\bibnamefont {Blank}}, \
  and\ \bibinfo {author} {\bibfnamefont {H.}~\bibnamefont {Hilgenkamp}},\
  }\href@noop {} {\bibfield  {journal} {\bibinfo  {journal} {Nature materials}\
  }\textbf {\bibinfo {volume} {6}},\ \bibinfo {pages} {493} (\bibinfo {year}
  {2007})}\BibitemShut {NoStop}%
\bibitem [{\citenamefont {Ueno}\ \emph {et~al.}(2008)\citenamefont {Ueno},
  \citenamefont {Nakamura}, \citenamefont {Shimotani}, \citenamefont {Ohtomo},
  \citenamefont {Kimura}, \citenamefont {Nojima}, \citenamefont {Aoki},
  \citenamefont {Iwasa},\ and\ \citenamefont {Kawasaki}}]{ueno2008electric}%
  \BibitemOpen
  \bibfield  {author} {\bibinfo {author} {\bibfnamefont {K.}~\bibnamefont
  {Ueno}}, \bibinfo {author} {\bibfnamefont {S.}~\bibnamefont {Nakamura}},
  \bibinfo {author} {\bibfnamefont {H.}~\bibnamefont {Shimotani}}, \bibinfo
  {author} {\bibfnamefont {A.}~\bibnamefont {Ohtomo}}, \bibinfo {author}
  {\bibfnamefont {N.}~\bibnamefont {Kimura}}, \bibinfo {author} {\bibfnamefont
  {T.}~\bibnamefont {Nojima}}, \bibinfo {author} {\bibfnamefont
  {H.}~\bibnamefont {Aoki}}, \bibinfo {author} {\bibfnamefont {Y.}~\bibnamefont
  {Iwasa}}, \ and\ \bibinfo {author} {\bibfnamefont {M.}~\bibnamefont
  {Kawasaki}},\ }\href@noop {} {\bibfield  {journal} {\bibinfo  {journal}
  {Nature materials}\ }\textbf {\bibinfo {volume} {7}},\ \bibinfo {pages} {855}
  (\bibinfo {year} {2008})}\BibitemShut {NoStop}%
\bibitem [{\citenamefont {Bert}\ \emph {et~al.}(2011)\citenamefont {Bert},
  \citenamefont {Kalisky}, \citenamefont {Bell}, \citenamefont {Kim},
  \citenamefont {Hikita}, \citenamefont {Hwang},\ and\ \citenamefont
  {Moler}}]{bert2011direct}%
  \BibitemOpen
  \bibfield  {author} {\bibinfo {author} {\bibfnamefont {J.~A.}\ \bibnamefont
  {Bert}}, \bibinfo {author} {\bibfnamefont {B.}~\bibnamefont {Kalisky}},
  \bibinfo {author} {\bibfnamefont {C.}~\bibnamefont {Bell}}, \bibinfo {author}
  {\bibfnamefont {M.}~\bibnamefont {Kim}}, \bibinfo {author} {\bibfnamefont
  {Y.}~\bibnamefont {Hikita}}, \bibinfo {author} {\bibfnamefont {H.~Y.}\
  \bibnamefont {Hwang}}, \ and\ \bibinfo {author} {\bibfnamefont {K.~A.}\
  \bibnamefont {Moler}},\ }\href@noop {} {\bibfield  {journal} {\bibinfo
  {journal} {Nature physics}\ }\textbf {\bibinfo {volume} {7}},\ \bibinfo
  {pages} {767} (\bibinfo {year} {2011})}\BibitemShut {NoStop}%
\bibitem [{\citenamefont {Nakagawa}\ \emph {et~al.}(2006)\citenamefont
  {Nakagawa}, \citenamefont {Hwang},\ and\ \citenamefont
  {Muller}}]{nakagawa2006some}%
  \BibitemOpen
  \bibfield  {author} {\bibinfo {author} {\bibfnamefont {N.}~\bibnamefont
  {Nakagawa}}, \bibinfo {author} {\bibfnamefont {H.~Y.}\ \bibnamefont {Hwang}},
  \ and\ \bibinfo {author} {\bibfnamefont {D.~A.}\ \bibnamefont {Muller}},\
  }\href@noop {} {\bibfield  {journal} {\bibinfo  {journal} {Nature materials}\
  }\textbf {\bibinfo {volume} {5}},\ \bibinfo {pages} {204} (\bibinfo {year}
  {2006})}\BibitemShut {NoStop}%
\bibitem [{\citenamefont {Pentcheva}\ and\ \citenamefont
  {Pickett}(2009)}]{pentcheva2009avoiding}%
  \BibitemOpen
  \bibfield  {author} {\bibinfo {author} {\bibfnamefont {R.}~\bibnamefont
  {Pentcheva}}\ and\ \bibinfo {author} {\bibfnamefont {W.~E.}\ \bibnamefont
  {Pickett}},\ }\href@noop {} {\bibfield  {journal} {\bibinfo  {journal}
  {Physical review letters}\ }\textbf {\bibinfo {volume} {102}},\ \bibinfo
  {pages} {107602} (\bibinfo {year} {2009})}\BibitemShut {NoStop}%
\bibitem [{\citenamefont {Cen}\ \emph {et~al.}(2008)\citenamefont {Cen},
  \citenamefont {Thiel}, \citenamefont {Hammerl}, \citenamefont {Schneider},
  \citenamefont {Andersen}, \citenamefont {Hellberg}, \citenamefont
  {Mannhart},\ and\ \citenamefont {Levy}}]{cen2008nanoscale}%
  \BibitemOpen
  \bibfield  {author} {\bibinfo {author} {\bibfnamefont {C.}~\bibnamefont
  {Cen}}, \bibinfo {author} {\bibfnamefont {S.}~\bibnamefont {Thiel}}, \bibinfo
  {author} {\bibfnamefont {G.}~\bibnamefont {Hammerl}}, \bibinfo {author}
  {\bibfnamefont {C.}~\bibnamefont {Schneider}}, \bibinfo {author}
  {\bibfnamefont {K.}~\bibnamefont {Andersen}}, \bibinfo {author}
  {\bibfnamefont {C.}~\bibnamefont {Hellberg}}, \bibinfo {author}
  {\bibfnamefont {J.}~\bibnamefont {Mannhart}}, \ and\ \bibinfo {author}
  {\bibfnamefont {J.}~\bibnamefont {Levy}},\ }\href@noop {} {\bibfield
  {journal} {\bibinfo  {journal} {Nature materials}\ }\textbf {\bibinfo
  {volume} {7}},\ \bibinfo {pages} {298} (\bibinfo {year} {2008})}\BibitemShut
  {NoStop}%
\bibitem [{\citenamefont {Jackiw}\ and\ \citenamefont
  {Rebbi}(1976)}]{jackiw1976solitons}%
  \BibitemOpen
  \bibfield  {author} {\bibinfo {author} {\bibfnamefont {R.}~\bibnamefont
  {Jackiw}}\ and\ \bibinfo {author} {\bibfnamefont {C.}~\bibnamefont {Rebbi}},\
  }\href@noop {} {\bibfield  {journal} {\bibinfo  {journal} {Physical Review
  D}\ }\textbf {\bibinfo {volume} {13}},\ \bibinfo {pages} {3398} (\bibinfo
  {year} {1976})}\BibitemShut {NoStop}%
\bibitem [{\citenamefont {Su}\ \emph {et~al.}(1979)\citenamefont {Su},
  \citenamefont {Schrieffer},\ and\ \citenamefont {Heeger}}]{su1979solitons}%
  \BibitemOpen
  \bibfield  {author} {\bibinfo {author} {\bibfnamefont {W.}~\bibnamefont
  {Su}}, \bibinfo {author} {\bibfnamefont {J.}~\bibnamefont {Schrieffer}}, \
  and\ \bibinfo {author} {\bibfnamefont {A.~J.}\ \bibnamefont {Heeger}},\
  }\href@noop {} {\bibfield  {journal} {\bibinfo  {journal} {Physical Review
  Letters}\ }\textbf {\bibinfo {volume} {42}},\ \bibinfo {pages} {1698}
  (\bibinfo {year} {1979})}\BibitemShut {NoStop}%
\bibitem [{\citenamefont {Rice}(1979)}]{rice1979charged}%
  \BibitemOpen
  \bibfield  {author} {\bibinfo {author} {\bibfnamefont {M.~J.}\ \bibnamefont
  {Rice}},\ }\href@noop {} {\bibfield  {journal} {\bibinfo  {journal} {Physics
  Letters A}\ }\textbf {\bibinfo {volume} {71}},\ \bibinfo {pages} {152}
  (\bibinfo {year} {1979})}\BibitemShut {NoStop}%
\bibitem [{\citenamefont {Su}\ \emph {et~al.}(1980)\citenamefont {Su},
  \citenamefont {Schrieffer},\ and\ \citenamefont {Heeger}}]{su1980soliton}%
  \BibitemOpen
  \bibfield  {author} {\bibinfo {author} {\bibfnamefont {W.-P.}\ \bibnamefont
  {Su}}, \bibinfo {author} {\bibfnamefont {J.}~\bibnamefont {Schrieffer}}, \
  and\ \bibinfo {author} {\bibfnamefont {A.}~\bibnamefont {Heeger}},\
  }\href@noop {} {\bibfield  {journal} {\bibinfo  {journal} {Physical Review
  B}\ }\textbf {\bibinfo {volume} {22}},\ \bibinfo {pages} {2099} (\bibinfo
  {year} {1980})}\BibitemShut {NoStop}%
\bibitem [{\citenamefont {Rao}\ \emph {et~al.}(2008)\citenamefont {Rao},
  \citenamefont {Sahu},\ and\ \citenamefont {Panigrahi}}]{rao2008fermion}%
  \BibitemOpen
  \bibfield  {author} {\bibinfo {author} {\bibfnamefont {K.}~\bibnamefont
  {Rao}}, \bibinfo {author} {\bibfnamefont {N.}~\bibnamefont {Sahu}}, \ and\
  \bibinfo {author} {\bibfnamefont {P.~K.}\ \bibnamefont {Panigrahi}},\
  }\href@noop {} {\bibfield  {journal} {\bibinfo  {journal} {Resonance}\
  }\textbf {\bibinfo {volume} {13}},\ \bibinfo {pages} {738} (\bibinfo {year}
  {2008})}\BibitemShut {NoStop}%
\bibitem [{\citenamefont {Jackiw}\ and\ \citenamefont
  {Schrieffer}(1981)}]{jackiw1981solitons}%
  \BibitemOpen
  \bibfield  {author} {\bibinfo {author} {\bibfnamefont {R.}~\bibnamefont
  {Jackiw}}\ and\ \bibinfo {author} {\bibfnamefont {J.~R.}\ \bibnamefont
  {Schrieffer}},\ }\href@noop {} {\bibfield  {journal} {\bibinfo  {journal}
  {Nuclear Physics B}\ }\textbf {\bibinfo {volume} {190}},\ \bibinfo {pages}
  {253} (\bibinfo {year} {1981})}\BibitemShut {NoStop}%
\bibitem [{\citenamefont {Bell}\ and\ \citenamefont
  {Rajaraman}(1983)}]{bell1983states}%
  \BibitemOpen
  \bibfield  {author} {\bibinfo {author} {\bibfnamefont {J.~S.}\ \bibnamefont
  {Bell}}\ and\ \bibinfo {author} {\bibfnamefont {R.}~\bibnamefont
  {Rajaraman}},\ }\href@noop {} {\bibfield  {journal} {\bibinfo  {journal}
  {Nuclear Physics B}\ }\textbf {\bibinfo {volume} {220}},\ \bibinfo {pages}
  {1} (\bibinfo {year} {1983})}\BibitemShut {NoStop}%
\bibitem [{\citenamefont {Cantoni}\ \emph {et~al.}(2012)\citenamefont
  {Cantoni}, \citenamefont {Gazquez}, \citenamefont {Miletto~Granozio},
  \citenamefont {Oxley}, \citenamefont {Varela}, \citenamefont {Lupini},
  \citenamefont {Pennycook}, \citenamefont {Aruta}, \citenamefont {di~Uccio},
  \citenamefont {Perna} \emph {et~al.}}]{cantoni2012electron}%
  \BibitemOpen
  \bibfield  {author} {\bibinfo {author} {\bibfnamefont {C.}~\bibnamefont
  {Cantoni}}, \bibinfo {author} {\bibfnamefont {J.}~\bibnamefont {Gazquez}},
  \bibinfo {author} {\bibfnamefont {F.}~\bibnamefont {Miletto~Granozio}},
  \bibinfo {author} {\bibfnamefont {M.~P.}\ \bibnamefont {Oxley}}, \bibinfo
  {author} {\bibfnamefont {M.}~\bibnamefont {Varela}}, \bibinfo {author}
  {\bibfnamefont {A.~R.}\ \bibnamefont {Lupini}}, \bibinfo {author}
  {\bibfnamefont {S.~J.}\ \bibnamefont {Pennycook}}, \bibinfo {author}
  {\bibfnamefont {C.}~\bibnamefont {Aruta}}, \bibinfo {author} {\bibfnamefont
  {U.~S.}\ \bibnamefont {di~Uccio}}, \bibinfo {author} {\bibfnamefont
  {P.}~\bibnamefont {Perna}},  \emph {et~al.},\ }\href@noop {} {\bibfield
  {journal} {\bibinfo  {journal} {Advanced Materials}\ }\textbf {\bibinfo
  {volume} {24}},\ \bibinfo {pages} {3952} (\bibinfo {year}
  {2012})}\BibitemShut {NoStop}%
\bibitem [{\citenamefont {Savoia}\ \emph {et~al.}(2009)\citenamefont {Savoia},
  \citenamefont {Paparo}, \citenamefont {Perna}, \citenamefont {Ristic},
  \citenamefont {Salluzzo}, \citenamefont {Granozio}, \citenamefont {di~Uccio},
  \citenamefont {Richter}, \citenamefont {Thiel}, \citenamefont {Mannhart}
  \emph {et~al.}}]{savoia2009polar}%
  \BibitemOpen
  \bibfield  {author} {\bibinfo {author} {\bibfnamefont {A.}~\bibnamefont
  {Savoia}}, \bibinfo {author} {\bibfnamefont {D.}~\bibnamefont {Paparo}},
  \bibinfo {author} {\bibfnamefont {P.}~\bibnamefont {Perna}}, \bibinfo
  {author} {\bibfnamefont {Z.}~\bibnamefont {Ristic}}, \bibinfo {author}
  {\bibfnamefont {M.}~\bibnamefont {Salluzzo}}, \bibinfo {author}
  {\bibfnamefont {F.~M.}\ \bibnamefont {Granozio}}, \bibinfo {author}
  {\bibfnamefont {U.~S.}\ \bibnamefont {di~Uccio}}, \bibinfo {author}
  {\bibfnamefont {C.}~\bibnamefont {Richter}}, \bibinfo {author} {\bibfnamefont
  {S.}~\bibnamefont {Thiel}}, \bibinfo {author} {\bibfnamefont
  {J.}~\bibnamefont {Mannhart}},  \emph {et~al.},\ }\href@noop {} {\bibfield
  {journal} {\bibinfo  {journal} {Physical Review B}\ }\textbf {\bibinfo
  {volume} {80}},\ \bibinfo {pages} {075110} (\bibinfo {year}
  {2009})}\BibitemShut {NoStop}%
\bibitem [{\citenamefont {Asmara}\ \emph {et~al.}(2014)\citenamefont {Asmara},
  \citenamefont {Annadi}, \citenamefont {Santoso}, \citenamefont {Gogoi},
  \citenamefont {Kotlov}, \citenamefont {Omer}, \citenamefont {Motapothula},
  \citenamefont {Breese}, \citenamefont {R{\"u}bhausen}, \citenamefont
  {Venkatesan} \emph {et~al.}}]{asmara2014mechanisms}%
  \BibitemOpen
  \bibfield  {author} {\bibinfo {author} {\bibfnamefont {T.}~\bibnamefont
  {Asmara}}, \bibinfo {author} {\bibfnamefont {A.}~\bibnamefont {Annadi}},
  \bibinfo {author} {\bibfnamefont {I.}~\bibnamefont {Santoso}}, \bibinfo
  {author} {\bibfnamefont {P.}~\bibnamefont {Gogoi}}, \bibinfo {author}
  {\bibfnamefont {A.}~\bibnamefont {Kotlov}}, \bibinfo {author} {\bibfnamefont
  {H.}~\bibnamefont {Omer}}, \bibinfo {author} {\bibfnamefont {M.}~\bibnamefont
  {Motapothula}}, \bibinfo {author} {\bibfnamefont {M.}~\bibnamefont {Breese}},
  \bibinfo {author} {\bibfnamefont {M.}~\bibnamefont {R{\"u}bhausen}}, \bibinfo
  {author} {\bibfnamefont {T.}~\bibnamefont {Venkatesan}},  \emph {et~al.},\
  }\href@noop {} {\bibfield  {journal} {\bibinfo  {journal} {Nature
  communications}\ }\textbf {\bibinfo {volume} {5}} (\bibinfo {year}
  {2014})}\BibitemShut {NoStop}%
\bibitem [{\citenamefont {Pauli}\ \emph {et~al.}(2011)\citenamefont {Pauli},
  \citenamefont {Leake}, \citenamefont {Delley}, \citenamefont {Bj{\"o}rck},
  \citenamefont {Schneider}, \citenamefont {Schlep{\"u}tz}, \citenamefont
  {Martoccia}, \citenamefont {Paetel}, \citenamefont {Mannhart},\ and\
  \citenamefont {Willmott}}]{pauli2011evolution}%
  \BibitemOpen
  \bibfield  {author} {\bibinfo {author} {\bibfnamefont {S.}~\bibnamefont
  {Pauli}}, \bibinfo {author} {\bibfnamefont {S.}~\bibnamefont {Leake}},
  \bibinfo {author} {\bibfnamefont {B.}~\bibnamefont {Delley}}, \bibinfo
  {author} {\bibfnamefont {M.}~\bibnamefont {Bj{\"o}rck}}, \bibinfo {author}
  {\bibfnamefont {C.}~\bibnamefont {Schneider}}, \bibinfo {author}
  {\bibfnamefont {C.}~\bibnamefont {Schlep{\"u}tz}}, \bibinfo {author}
  {\bibfnamefont {D.}~\bibnamefont {Martoccia}}, \bibinfo {author}
  {\bibfnamefont {S.}~\bibnamefont {Paetel}}, \bibinfo {author} {\bibfnamefont
  {J.}~\bibnamefont {Mannhart}}, \ and\ \bibinfo {author} {\bibfnamefont
  {P.}~\bibnamefont {Willmott}},\ }\href@noop {} {\bibfield  {journal}
  {\bibinfo  {journal} {Physical review letters}\ }\textbf {\bibinfo {volume}
  {106}},\ \bibinfo {pages} {036101} (\bibinfo {year} {2011})}\BibitemShut
  {NoStop}%
\bibitem [{\citenamefont {Salluzzo}\ \emph {et~al.}(2013)\citenamefont
  {Salluzzo}, \citenamefont {Gariglio}, \citenamefont {Torrelles},
  \citenamefont {Ristic}, \citenamefont {Di~Capua}, \citenamefont {Drnec},
  \citenamefont {Sala}, \citenamefont {Ghiringhelli}, \citenamefont {Felici},\
  and\ \citenamefont {Brookes}}]{salluzzo2013structural}%
  \BibitemOpen
  \bibfield  {author} {\bibinfo {author} {\bibfnamefont {M.}~\bibnamefont
  {Salluzzo}}, \bibinfo {author} {\bibfnamefont {S.}~\bibnamefont {Gariglio}},
  \bibinfo {author} {\bibfnamefont {X.}~\bibnamefont {Torrelles}}, \bibinfo
  {author} {\bibfnamefont {Z.}~\bibnamefont {Ristic}}, \bibinfo {author}
  {\bibfnamefont {R.}~\bibnamefont {Di~Capua}}, \bibinfo {author}
  {\bibfnamefont {J.}~\bibnamefont {Drnec}}, \bibinfo {author} {\bibfnamefont
  {M.~M.}\ \bibnamefont {Sala}}, \bibinfo {author} {\bibfnamefont
  {G.}~\bibnamefont {Ghiringhelli}}, \bibinfo {author} {\bibfnamefont
  {R.}~\bibnamefont {Felici}}, \ and\ \bibinfo {author} {\bibfnamefont
  {N.}~\bibnamefont {Brookes}},\ }\href@noop {} {\bibfield  {journal} {\bibinfo
   {journal} {Advanced Materials}\ }\textbf {\bibinfo {volume} {25}},\ \bibinfo
  {pages} {2333} (\bibinfo {year} {2013})}\BibitemShut {NoStop}%
\end{thebibliography}%
\end{document}